\pdfoutput=1
\PassOptionsToPackage{obeyspaces}{url}
\documentclass[twocolumn,prd,a4paper,final,superscriptaddress,longbibliography,nofootinbib,floatfix]{revtex4-1}

\usepackage{amsmath,amssymb}
\usepackage[dvipdfmx]{graphicx}
\usepackage{comment}
\usepackage{bm}
\usepackage[colorlinks=true,allcolors=blue]{hyperref}
\usepackage{color}

\newcommand{\Eprint}[2]{\href{#1}{\urlstyle{same}\nolinkurl{#2}}}

\newcommand{\pd}{\partial}
\newcommand{\Tr}{\qopname\relax o{Tr}}

\newcommand{\n}{\mathfrak{n}}
\newcommand{\rmL}{\mathrm{L}}

\newcommand{\cl}{\mathrm{cl}}

\def\ket#1{\left|{#1}\right>}

\newcommand{\braket}[2]{\left< {#1} | {#2}\right>}

\usepackage{multirow}

\begin{document}

\title{
Glueballonia as Hopfions
}
	
\date{\today}

\author{ Yuki Amari}
\affiliation{Department of Physics $\&$ Research and Education Center for Natural Sciences, Keio University, 4-1-1 Hiyoshi, Yokohama, Kanagawa 223-8521, Japan}

\author{Muneto Nitta}
\affiliation{Department of Physics $\&$ Research and Education Center for Natural Sciences, Keio University, 4-1-1 Hiyoshi, Yokohama, Kanagawa 223-8521, Japan}
\affiliation{International Institute for Sustainability with Knotted Chiral Meta Matter (WPI-SKCM$^2$), Hiroshima University, 1-3-1 Kagamiyama, Higashi-Hiroshima, Hiroshima 739-8526, Japan}

\author{Chihiro Sasaki}
\affiliation{Institute of Theoretical Physics, University of Wroclaw, plac Maksa Borna 9, PL-50204 Wroclaw,
Poland}
\affiliation{International Institute for Sustainability with Knotted Chiral Meta Matter (WPI-SKCM$^2$), Hiroshima University, 1-3-1 Kagamiyama, Higashi-Hiroshima, Hiroshima 739-8526, Japan}

\author{Kenta Shigaki}
\affiliation{Physics Program, Hiroshima University, 1-3-1 Kagamiyama, Higashi-Hiroshima, Hiroshima 739-8526, Japan}
\affiliation{International Institute for Sustainability with Knotted Chiral Meta Matter (WPI-SKCM$^2$), Hiroshima University, 1-3-1 Kagamiyama, Higashi-Hiroshima, Hiroshima 739-8526, Japan}

\author{Satoshi Yano}
\affiliation{Physics Program, Hiroshima University, 1-3-1 Kagamiyama, Higashi-Hiroshima, Hiroshima 739-8526, Japan}
\affiliation{International Institute for Sustainability with Knotted Chiral Meta Matter (WPI-SKCM$^2$), Hiroshima University, 1-3-1 Kagamiyama, Higashi-Hiroshima, Hiroshima 739-8526, Japan}

\author{Shigehiro Yasui}
\affiliation{Nishogakusha University, 6-16, Sanbancho, Chiyoda, Tokyo 102-8336, Japan}
\affiliation{International Institute for Sustainability with Knotted Chiral Meta Matter (WPI-SKCM$^2$), Hiroshima University, 1-3-1 Kagamiyama, Higashi-Hiroshima, Hiroshima 739-8526, Japan}
\affiliation{Department of Physics $\&$ Research and Education Center for Natural Sciences, Keio University, 4-1-1 Hiyoshi, Yokohama, Kanagawa 223-8521, Japan}

\vspace{.5in}

\begin{abstract}
We work out the Hopfion description of glueballs 
by 
inclusively comparing the energy spectra obtained by quantizing Hopfions with experimental data and lattice QCD. 
Identifying 
a Hopfion carrying 
a unit topological charge as
$f_0(1500)$,
the Hopfions with the topological charge two 
are classified as glueballonia, 
i.e.,~two glueballs are bound together.
We find a
 tightly and a loosely bound glueballonia  complying with 
$f_0 (2470)$ and 
a novel scalar particle carrying the mass 
around 2814 MeV, respectively, 
and calculate their binding energies.
By the rigid body quantization of Hopfions, we predict a characteristic multiplet structure of 
tensor glueball states. 
Some of them
 are missing in the current experimental data and can be verified in future measurements.
\end{abstract} 

\maketitle


\section{Introduction}

One of the most challenging problems in modern science is {\it color confinement} and {\it the origin of mass}   
in quantum chromodynamics (QCD) 
describing the strong interaction, 
one of the fundamental forces in nature, 
as 
it is one of the millennium problems of
mathematics~\cite{millennium}.
Quarks are not observed in nature
and instead are confined into a form of color singlet particles called 
hadrons, i.e.,~
baryons (such as protons and neutrons) consisting of three quarks and mesons consisting of a pair of quark and antiquark. 
Recently, apart from these standard hadrons,
various candidates of
exotic hadrons
with different quark compositions
have been observed 
in several experiments. 
Among them, gluesballs are 
one of the most intriguing particles,
because they are made purely of the gauge fields, i.e., gluons, 
without any quarks, hence classically massless but acquire their masses quantum-mechanically 
in pure Yang-Mills theory/QCD
~\cite{Fritzsch1975,Jaffe:1985qp,Close1991} (see, e.g., Refs.~\cite{Klempt:2007cp,Mathieu:2008me,Crede:2008vw,Greensite:2011zz,Ochs:2013gi,Shepherd:2016dni,Llanes-Estrada:2021evz,Chen:2022asf,Vadacchino:2023vnc,Morningstar:2024vjk} for reviews).
Thus, understanding their characteristics will offer us a promising guide to reveal 
the mechanism of color confinement and dynamical mass generation.

Besides the first-principle approach,
Lattice QCD (LQCD) \cite{Johnson:1997ap,Michael:1988jr,Bali:1993fb,Sexton:1995kd,Morningstar:1999rf,Johnson:2000qz,Liu:2001wqa,Meyer:2004jc,Chen:2005mg,Loan:2006gm,Gregory:2012hu,Brett:2019tzr,Athenodorou:2020ani,Chen:2021dvn,Sakai:2022zdc,Morningstar:2024vjk}, 
the properties of glueballs, e.g., masses and decays, have also been studied in a variety of phenomenological models, 
such as a flux-tube model
~\cite{Isgur:1984bm,Kuti:1998rh},
bag model~\cite{Jaffe:1975fd},
constituent gluon model
~\cite{Barnes:1981ac,Cornwall:1982zn},
and 
in
holographic QCD~\cite{Witten:1998zw,Sakai:2004cn}
based on string theory~\cite{Csaki:1998qr,deMelloKoch:1998vqw,Brower:2000rp,Amador:2004pz,Colangelo:2007pt,Hashimoto:2007ze,Forkel:2007ru,Miranda:2009qp,Brunner:2015oqa,Brunner:2015yha,RodriguesFilho:2020rae}.
It has recently been reported that a glueball candidate, $X(2370)$, was observed in high-energy experiments by 
BESIII~\cite{BESIII:2023wfi,
She:2024ewy,Liu:2024anq} 
after some earlier attempts~\cite{Crede:2009}.
The observed mass
around $2.3$ - $2.6$ GeV 
is comparable with 
the lattice QCD prediction 
and that of  
holographic QCD\@.

However, identifying glueballs in experiments remains tremendously difficult.
This can be overcome if one could know their internal structures, 
such as their sizes, shapes, and energy density distributions as well as their masses,
productions and decays.
In contrast, 
these problems are accessible in 
the Hopfion approach 
\cite{Faddeev:1996zj,Faddeev:1998eq,Faddeev:2003aw,Kondo:2006sa} 
in which 
glueballs are semiclasically described 
as Hopfions 
or knot solitons, i.e.,~topological solitons composed of gluons,  
in parallel to the Skyrme model \cite{Skyrme:1961vq,Skyrme:1962vh} in which nucleons are described as 
Skyrmions, i.e.,~topological solitons composed of pions.
Apart from 
QCD, Hopfions have been extensively studied in various condensed matter systems 
such as 
${}^3$He superfluid~\cite{Volovik1977JETP,Volovik:2003fe}, 
superconductors~\cite{Babaev:2001zy,Rybakov:2018ktd},
spinor Bose-Einstein condensates \cite{Kawaguchi:2008xi,Hall:2016,Ollikainen:2019dyh},  
liquid crystals and colloids~
\cite{PhysRevLett.110.237801,Ackerman:2015,Ackerman:2017,Ackerman:2017b,Tai:2018,Tai:2019,Smalyukh:2020zin,Smalyukh:2022}, 
magnets~\cite{Sutcliffe:2017aro,Sutcliffe:2018vcb,Kent:2020jvm}, 
and pure electromagnetism~\cite{Trautman:1977im,Ranada:1989wc,Kedia:2013bw,Arrayas:2017sfq}.

In this Letter, we 
work out 
glueballs as Hopfions
and confront the obtained spectra, for the first time, with currently available experimental data. 
While the Hopfion with a unit topological charge is naturally identified as the scalar particle $f_0(1500)$, 
the Hopfions carrying the topological charge two 
are associated with
$f_0 (2470)$ with an unexpectedly accurate mass and 
a new scalar particle with mass around 2814 MeV.
We find
that they are 
glueballonia~\cite{Giacosa:2021brl}, i.e.,~two glueballs 
tightly and loosely bound together, respectively,
and the latter is an excited state of the former. 
Their binding energies are also estimated.
We further predict a distinctive multiplet-structure of 
tensor glueballs of angular momentum two; the lowest and second lowest scalar glueballs, $f_0(1500)$ and $f_0 (2470)$, 
can be quantum mechanically excited to triplet and doublet tensor glueballs, respectively.  
These novel relations among scalar and tensor glueballs are
the manifestation
of our Hopfion approach.

\section{Quantization of Hopfions}

It was proposed in Refs.~\cite{Faddeev:1998eq,Shabanov:1999xy,Shabanov:1999uv} that the $\mathrm{SU}(2)$
pure Yang-Mills theory reduces to 
an $\mathrm{O}(3)$ nonlinear sigma model 
complemented by a 
four derivative term, 
called the Skyrme-Faddeev model~
\footnote{Note that there is also another derivation 
accompanied with a potential term 
\cite{Faddeev:2001dda}. 
There are also more rigorous derivations 
based on the renormalization group analysis \cite{Gies:2001hk}
and the heat karnel expansion \cite{Langmann:1999nn}. 
In addition, generalizations of the proposal to the $\mathrm{SU}(N)$ have been done in Refs.~\cite{Faddeev:1998yz,Faddeev:1999cj,Kondo:2008xa,Evslin:2011ti,Kondo:2014sta}. Hopfions and a string junction in the $\mathrm{SU}(3)$ generalized version of the Skyrme-Faddeev model have been discussed in Refs.~\cite{Amari:2018gbq,Amari:2024vsf}.
There is also a criticism of the proposal \cite{Evslin:2010sb,Niemi:2010ms}.
}. 
The Lagrangian density
is given by~\cite{Faddeev:1996zj,Faddeev:1998eq}
\begin{equation}
    {\cal L}=\frac{\kappa^2}{4}\Tr\left(\pd_\mu \n \pd^\mu \n \right)+\frac{1}{32e^2}\Tr\left[\pd_\mu \n,\pd_\nu \n \right]^2,
    \label{eq:Lag-density}
\end{equation}
with $\n = \bm{\tau}\cdot\bm{n}$ where $\bm{n}=(n_1,n_2,n_3)$ is a three-component unit vector 
parameterizing the two-dimensional sphere 
$S^2$,
and $\bm{\tau}$ is a vector whose components are Pauli matrices. The coupling constant $\kappa$ has the dimension of mass, and $e$ is dimensionless.

This model admits Hopfions as finite energy solutions with a non-zero topological invariant $Q$, called Hopf charge,
taking a value in an integer 
$\mathbb{Z}$
associated with the third homotopy group $\pi_3(S^2)\simeq\mathbb{Z}$ 
\cite{Gladikowski:1996mb,Faddeev:1996zj,Battye:1998pe,Hietarinta:1998kt,Hietarinta:2000ci,Sutcliffe:2007ui}. 
Here we concentrate on low-lying glueballs, and therefore, we only consider Hopfions with topological charge $Q=1,2$.
It is well known that these solutions 
exhibit an axial symmetry~\cite{Gladikowski:1996mb,Battye:1998pe}.
Such an axially symmetric Hopfion can be represented as a torus, where each normal slice is a two-dimensional soliton defined by a map  
$\bm{n}: S^2 \to S^2$,  
called baby Skyrmion.  
The two-dimensional soliton is classified by an integer $ m \in \pi_2(S^2) $, 
the second homotopy group.
In addition, as it travels once around the circle, the internal phase of the two-dimensional soliton rotates by $ 2\pi \ell $ with $ \ell \in \mathbb{Z} $. For such configurations, the Hopf charge is 
given by $Q=\ell m$~\footnote{
This construction gives an interpretation of Hopfions as torus knots
\cite{Kobayashi:2013xoa}.
}.

By quantizing a Hopfion as a rigid body 
in the collective coordinate quantization
~\cite{Su:2001zw,Krusch:2005bn,Kondo:2006sa,Acus:2012st}, 
one can construct energy spectra of Hopfions 
confronted with LQCD data~\cite{Kondo:2006sa}, similarly toquantized Skyrmions that can describe baryons and nuclei~\cite{Adkins:1983ya,Braaten:1988cc}.
The quantization is implemented by promoting parameters associated with symmetries of the static energy, and then by quantizing the dynamical system according to the canonical method.
Specifically, we utilize a dynamical ansatz of the form
\begin{equation}
    \hat{\n}(\bm{x},t)\equiv A(t)\n(R(B(t))\bm{x})A(t)^\dagger,
    \label{eq:dynamical_ansatz_main}
\end{equation}
where $A$ is an $\mathrm{SU}(2)$ matrix associated with the isospatial rotation symmetry and 
$R_{jk}(B)=\frac{1}{2}\Tr\left(\tau_jB\tau_k B^\dagger \right)$ with an $\mathrm{SU}(2)$ matrix $B$ associates to the spatial rotation symmetry~\footnote{We ignored the translational degrees of freedom by quantizing Hopfions in its rest frame.}.
Substituting the ansatz \eqref{eq:dynamical_ansatz_main} into the Lagrangian 
density 
(\ref{eq:Lag-density})
and 
integrating over three-dimensional space, 
one finds Lagrangian
\begin{equation}
    L \,[\hat{\n}]=\frac{1}{2}a_j U_{jk} a_k - a_j W_{jk}b_k + \frac{1}{2}b_j V_{jk} b_k -M_\cl,\label{eq:Lagrangian}
\end{equation}
where $U_{jk},~W_{jk}$ and $V_{jk}$ are the inertial tensors, and $M_\cl$ is the static mass \footnote{
See Supplemental Material~(Sec.~I) for the detailed information.
}.
The valuables $a_j$ and $b_j$ denote angular velocities of isospatial and spatial rotation, respectively, defined as $a_k=-i\Tr(\tau_kA^\dagger \pd_t A)$
and
$b_k=i\Tr(\tau_k \pd_t BB^\dagger)$.
By the Legendre transformation, one obtains the Hamiltonian, and thanks to the axial symmetry, it reduces to that of a symmetric top.

As the result, the energy eigenvalue of quantized Hopfions is given by
\begin{equation}
    E=M_\cl + \frac{J(J+1)}{2V_{11}}+\left(\frac{1}{U_{33}}-\frac{\ell^2}{V_{11}} \right)\frac{K_3^2}{2},
    \label{eq:q-energy}
\end{equation}
where $J$ is the spin quantum number and $K_3$ is a quantum number analogous to the third component of isospin. Note that $K_3$ has no relation to the flavor symmetry for up and down quarks.
We can evaluate the energy by fixing the winding numbers $(\ell, m)$ and quantum numbers $(J,|K_3|)$ for a set of given coupling constants.

The basis for the eigenstate is the direct product states 
$\ket{J,J_3,L_3} \ket{I,I_3,K_3}$, where $|J_3|,|L_3|\leq J$ and $|I_3|,|K_3|\leq I$. 
Moreover, we have the constraint $L_3=\ell K_3$ because an isorotation around the $n_3$-axis and spatial rotation around the $x^3$-axis are identified as axial symmetric configurations. 
Following the spatial inversion property of the wave function, we define the parity quantum number as 
    $P=(-1)^J$.
Therefore, we find that the quantized Hopfions can be labeled by $\{(\ell,m),(J,|K_{3}|)\}$ with $|\ell K_3|\leq J$.

\section{Numerical Results}

\begin{figure}[!t]
    \centering
    \includegraphics[width=26em]{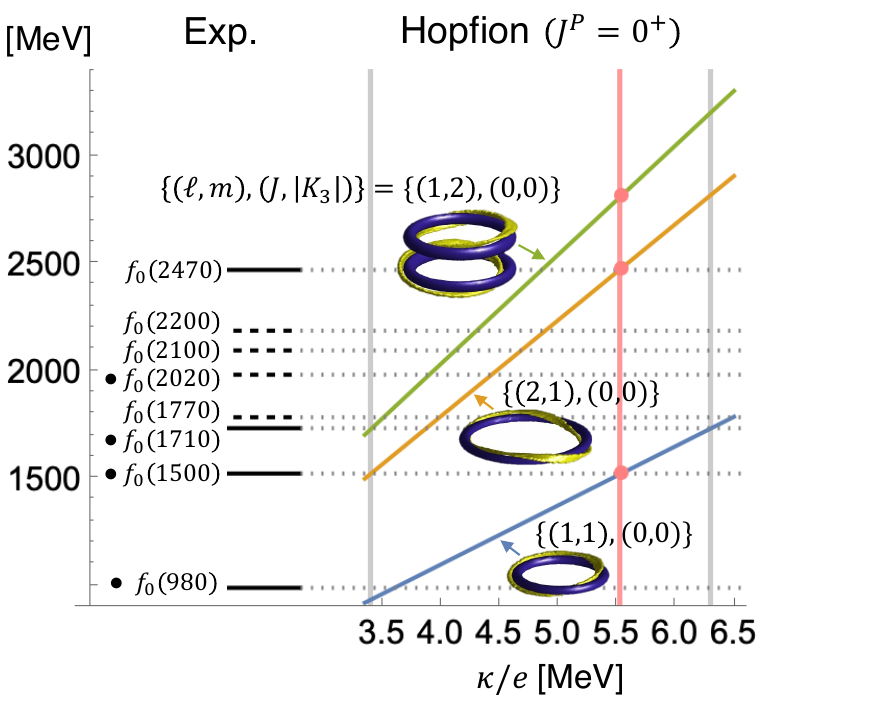}
    \caption{
    Energy spectra of the Hopfions with $J=0$. On the r.h.s.\ of the vertical axis, the energies of the Hopfion with the winding numbers $(\ell,m)=(1,1)$, $(2,1)$ and  $(1,2)$ are plotted as functions of $\kappa/e$. 
    The red vertical dotted line corresponds to 
    Eq.~(\ref{eq:k/e}) while 
    the gray vertical dotted  lines correspond to $\kappa/e = 3.40$ MeV and $6.31$ MeV
    discussed in Supplemental Material.
    The profile of each Hopfion is also plotted: the purple isosurface shows the field line with $n_3=-1$ representing the Hopfion's position, and the yellow shows $\bm{n}=(\sqrt{2\mu-\mu^2},0,-1+\mu)$ with $\mu=0.1$ representing the linking. On the l.h.s., the mass spectrum of light unflavored mesons with $J^{PC}=0^{++}$ in PDG~\cite{ParticleDataGroup:2022pth} is given.
    The solid (dashed) lines stand for particles with a decay width smaller (larger) than $150$ MeV, and the symbols $\bullet$ indicate established particles. 
    }
    \label{fig:f0}
\end{figure}

\begin{figure}[!t]
    \centering
    \includegraphics[width=26em]{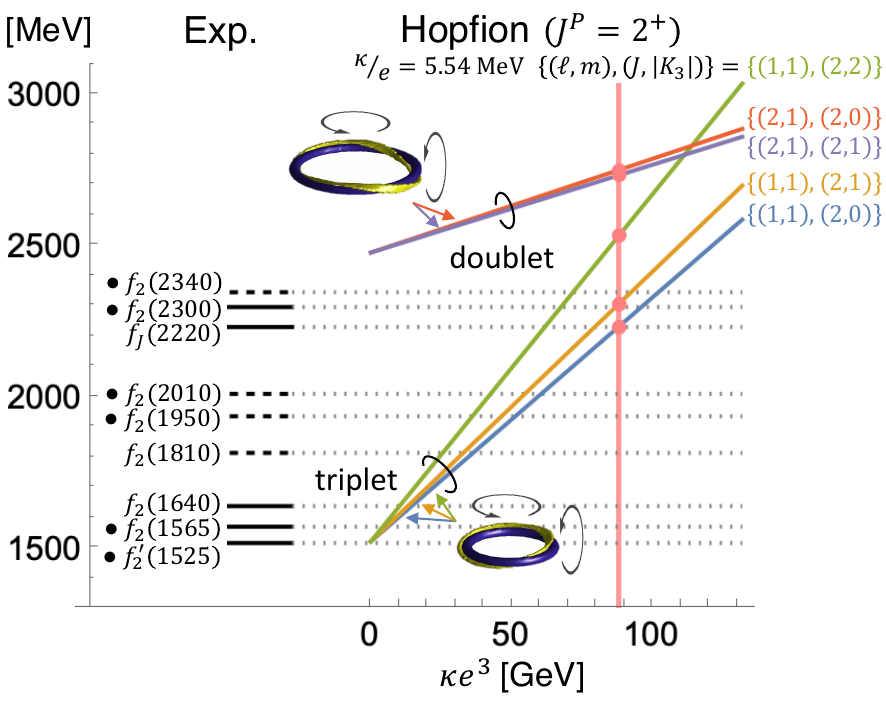}
    \caption{
    Energy spectra of the Hopfions with $J=2$
    for Eq.~(\ref{eq:k/e}).
    The Hopfion energies are plotted as functions of $\kappa e^3$ to the r.h.s of the vertical axis. On the l.h.s., the mass spectra of light unflavored mesons with $J^{PC}=2^{++}$ collected in PDG~\cite{ParticleDataGroup:2022pth} is given.  We also included $f_J(2220)$ with assuming $J=2$.
   The red vertical dotted line corresponds to $\kappa e^3 = 88.5$ GeV, where the energy of the first excited state $\{(\ell,m),(J,|K_3|)\} = \{(1,1),(2,0)\}$ equals to the mass of $f_J(2220)$ as a possible candidate of the tensor glueball.
    }
    \label{fig:f2}
\end{figure}

\begin{figure*}[!t]
    \centering
    \includegraphics[width=50em]{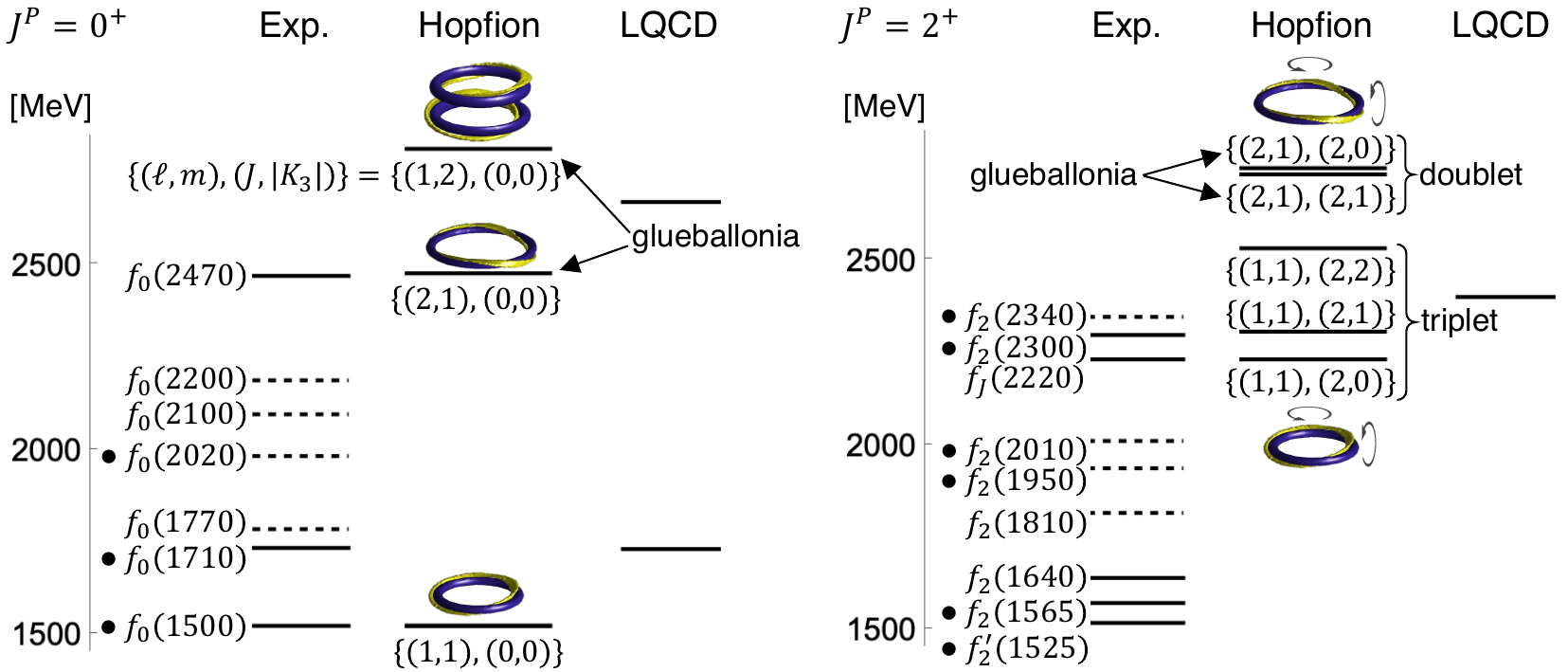}
    \caption{
    Comparison between the Hopfion energy spectra and the mass spectra of light unflavored mesons listed in PDG~\cite{ParticleDataGroup:2022pth} and glueballs predicted by LQCD~\cite{Morningstar:1999rf}. Left panel is for $0^{++}$ and right panel $2^{++}$. The symbol $\bullet$ indicates established particles. We used the values of the coupling constants in Eqs.~\eqref{eq:k/e} and \eqref{eq:ke3}. 
    } 
    \label{fig:spectrum}
\end{figure*}


\begin{table}[!t]
\centering
\caption{Hopfion energy eigenvalues ($E$) with the values of the coupling constants in Eqs.~\eqref{eq:k/e} and \eqref{eq:ke3}.
The last column shows the masses of scalar and tensor mesons, $f_{0}(1500)$, $f_{0}(2470)$, $f_{J=2}(2220)$, $f_{2}(2300)$ and $f_{2}(2340)$~\cite{ParticleDataGroup:2022pth}.
The mesons used for the fitting are indicated by asterisks.
All the states with $Q=2$ are candidates of glueballonia.
See the text.
}
\begin{tabular}{cc|cccc}
\hline\hline
$Q$ & $(\ell,m)$ & multiplet & $(J,|K_3|)$   & ~~$E$ [MeV] ~~ & Exp. [MeV] \\
\hline
1 & (1,1) & singlet & (0,0) & 1522 &  \,\,1500$^{\ast}$ \\
\cline{3-6}
  &       &         & (2,0) & 2231 &  \,\,2220$^{\ast}$ \\
  &       & triplet & (2,1) & 2306 &  2300 \\
  &       &         & (2,2) & 2531 &  2340 \\
      \hline
2 & (2,1) & singlet & (0,0) & 2477 &  2470 \\
\cline{3-6}
  &       & \multirow{2}{*}{doublet} & (2,0) & 2749 &  --- \\
  &       &         & (2,1) & 2732 &  --- \\
      \hline
2 & (1,2) & singlet & (0,0) & 2814 &  --- \\
\cline{3-6}
  &       &         & (2,0) & 3034 &  ---\\
  &       & triplet & (2,1) & 3108 &  ---\\
  &       &         & (2,2) & 3330 &  ---\\
\hline\hline
\end{tabular}
\label{table:hopfion_energy}
\end{table}

To evaluate the Hopfion energy spectrum, we
need
to fix the coupling constants. 
In the length scale $1/\kappa e$, the static mass is proportional to $\kappa/e$ and the quantum correction to $1/\kappa e^3$~\footnote{See Supplemental Material~(Sec.~II) for the numerical value of the static mass and inertia tensors used for 
the energy eigenvalue.}.
Here we set the parameters $\kappa/e$ and $\kappa e^3$ by identifying the Hopfions as light unflavored (isosinglet) mesons carrying spin-parity $J^{P}=0^{+}$ and $2^{+}$ listed by the Particle Data Group (PDG)~\cite{ParticleDataGroup:2022pth}.

Identifying
the $Q=1$
Hopfion with $\{(\ell,m),(J,|K_{3}|)\}=\{(1,1),(0,0)\}$
as the scalar meson 
$f_0(1500)$, which is one of the primary
candidates for the lowest-lying scalar glueball~\cite{Morningstar:1999rf}, we obtain
\begin{equation}
    \kappa/e = 
    5.54~\text{MeV} .
    \label{eq:k/e}
\end{equation}
Once the parameter $\kappa/e$ is fixed, 
the energy spectrum of the Hopfions with $J^{P}=0^{+}$ is fully determined as the classical solution, as shown in Fig.~\ref{fig:f0}.
There one observes
that 
the $Q=2$ Hopfion
carrying $\{(2,1),(0,0)\}$
with the value~\eqref{eq:k/e} acquires the mass 
extremely close to $f_0(2470)$, which is very narrow with the decay width, $75^{+14}_{-12}$ MeV, compared to its large mass~\cite{ParticleDataGroup:2022pth}.
Thus, the parameter set~\eqref{eq:k/e} is successful for reproducing the masses of $f_{0}(1500)$ and $f_{0}(2470)$ simultaneously.
Besides, we
find another $Q=2$
Hopfion with $\{(1,2),(0,0)\}$
with mass 2814 MeV, which has no corresponding state in the PDG list yet. 
This 
novel scalar meson is expected to be verified in future experiments.

With the given
$\kappa/e$,
the masses of Hopfions with $J^{P}=2^{+}$ can also be obtained.
They include quantum fluctuations beyond the classical solution.
Figure~\ref{fig:f2} shows the energy spectrum of the $J^{P}=2^{+}$ Hopfions with the value in~Eq.~\eqref{eq:k/e}.
The lowest state, the $Q=1$ Hopfion with $\{(1,1),(2,0)\}$, can naturally be identified as the narrowest tensor glueball, $f_J(2220)$, with the decay width $23^{+8}_{-7}$ MeV~\cite{ParticleDataGroup:2022pth}, assuming $J=2$.
This results in
\begin{equation}
    \kappa e^3=88.5~\text{GeV} ,
    \label{eq:ke3}
\end{equation}
leading to the full $2^{+}$ Hopfion energy spectrum. 
One finds that the energy of 
another $Q=1$ Hopfion
with $\{(1,1),(2,1)\}$
well coincides with the mass of $f_2(2300)$.
There is also the $Q=1$ Hopfion with $\{(1,1),(2,2)\}$ which may be comparable with $f_{2}(2340)$.
Thus, the Hopfion approach predicts the triplet structure, i.e.,
$f_{J=2}(2220)$, $f_2(2300)$ and $f_2(2340)$, and all of which are possible candidates of tensor glueballs 
in contrast to the singlet structures in $0^{+}$.
We also
find 
a doublet of $Q=2$ Hopfions with $\{(2,1),(2,1)\}$ and $\{(2,1),(2,0)\}$,
which is a genuine prediction of the Hopfion approach and should be experimentally assessed if they would serve as tensor glueballs. 

We summarize the
Hopfion energy spectra  in Table~\ref{table:hopfion_energy}.
Figure.~\ref{fig:spectrum} shows a direct comparison of these results with
the masses of scalar and tensor mesons
in PDG and
the glueball masses predicted in LQCD~\cite{Morningstar:1999rf}.
It should be emphasized that in the Hopfion approach the masses of
$f_{0}(2470)$, 
$f_{2}(2300)$ and $f_{2}(2340)$ 
are determined systematically without any fine-tuning.
One may consider other parameter settings, which however turns out to be unfavorable~\footnote{See Supplemental Material (Sec.~III) for detailed discussion.
}.

The Hopfion approach also provides the interpretation that the heavier scalar glueballs are bound states of the lightest glueballs, i.e.,~glueballonia~\cite{Giacosa:2021brl}.
The two $Q=2$ Hopfions 
with $(\ell,m)=(2,1)$ and $(1,2)$ can be identifed as a tightly and a loosely bound glueballonia 
of masses $2477$ MeV and $2814$ MeV, respectively,
possessing the binding energies of
$576$ MeV and $230$ MeV
calculated in this model.
Whereas the former is identified as the known state, $f_0(2470)$, 
the latter is a novel exotic particle.

\section{Summary}

We developed the Hopfion approach to describe low-lying scalar and tensor glueballs, confronted with the currently available data from experiments and LQCD as well as the QCD phenomenology.
Our model predicts that $f_0(2470)$ is a tightly bound glueballonium on top of a loosely bound one as a novel scalar state.
Thus, the Hopfion approach explains the unnaturally long lifetime of $f_0(2470)$, while an observation of the new glueballonium in future measurements would be a critical verification of this approach.
We also found several other Hopfions with $Q=1$ and $2$ corresponding to known and unknown glueballs, which are classified into the characteristic multiplet structures anchored to the underlying topology.

The current approach captures the peculiar features to glueballs complying with the existing data from measurements and lattice simulations, while this can be further refined:
We used
the rigid body quantization for tensor glueballs,
whereas the stationary isospinning solutions have been studied beyond the rigid body taking into account the deformation due to rotation~\cite{Harland:2013uk,Battye:2013xf}.
Including this effect may produce somecorrections to the spectra 
of tensor glueballs presented in this Letter.

Formation of the glueballonia from two 
glueballs may be investigated
in synergy with relativistic heavy-ion collisions 
and real time dynamics of Hopfions.
Quantum transitions between different 
glueball states should be studied in the Hopfion approach.
Modeling Hopfions coupled to dynamical quarks is also desired to
study decays of glueballs into quark-antiquark pairs to be compared with experimental data. 

\section{Acknowledgments}
YA is grateful for the kind hospitality at the Instituto de Física de São Carlos of Universidade de São Paulo (IFSC/USP).
This work is supported in part by 
 JSPS KAKENHI [Grants  No.~JP23KJ1881 (YA), No. JP22H01221 (MN)] and the WPI program ``Sustainability with Knotted Chiral Meta Matter (SKCM$^2$)'' at Hiroshima University.
The work of CS was supported partly by the Polish
National Science Centre (NCN) under OPUS Grant
No. 2022/45/B/ST2/01527.
The numerical computations in this paper were run on the ``GOVORUN" cluster supported by the LIT, JINR.
 
\bibliographystyle{apsrev4-1}
%


\newpage
\onecolumngrid     

\section*{Supplemental material}
\setcounter{equation}{0}
\renewcommand{\theequation}{S.\arabic{equation}}
\setcounter{figure}{0}
\renewcommand{\thefigure}{S\arabic{figure}}
\setcounter{table}{0}
\renewcommand{\thetable}{S\arabic{table}}

\section{I.~Explicit form of the classical mass and inertia tensors}
\label{sec:S1}

In this section, we describe the quantization of rotation modes 
of Hopfions.
\subsection{Semiclassical quantization}
The static energy of the model is invariant under the translation, spatial rotation and isospatial rotation, which is described by the transformation
\begin{equation}
   \n(\bm{x})\to \n(\bm{x}; \bm{X},A,B)\equiv A\n(R(B)(\bm{x}-\bm{X}))A^\dagger ,
   \label{eq:symmetry}
\end{equation}
where $\bm{X}$ denotes a translation, 
$A$ is an $SU(2)$ matrix associated with an isospatial rotation, and spatial rotation is represented by
\begin{equation}
    R_{jk}(B)=\frac{1}{2}\Tr\left(\tau_jB\tau_k B^\dagger \right) ,
\end{equation}
with an $SU(2)$ matrix $B$.
The degeneracy described in Eq.~\eqref{eq:symmetry} is lifted when the theory is quantized.
We apply a collective coordinate quantization \cite{Adkins:1983ya,Braaten:1988cc,Krusch:2005bn,Kondo:2006sa} implemented by promoting the parameters $\bm{X}, A$, and $B$ to the dynamical variable $\bm{X}(t), A(t)$, and $B(t)$, and then by quantizing the dynamical system according to the canonical method.
Therefore, we consider a dynamical ansatz of the form
\begin{equation}
    \hat{\n}(\bm{x},t)\equiv A(t)\n(R(B(t))\bm{x})A(t)^\dagger ,
    \label{eq:dynamical_ansatz}
\end{equation}
where we have ignored the translational degrees of freedom $\bm{X}(t)$, which means that we quantize a Hopfion in its rest frame.
Substituting the ansatz \eqref{eq:dynamical_ansatz} into the Lagrangian density in Eq.~(\ref{eq:Lag-density}) 
and integrating over three-dimensional space, 
one finds the effective Lagrangian 
\begin{equation}
    L[\hat{\n}]=T-M_\cl ,
\end{equation}
where $M_\cl$ stands for the static energy of the background field $\n(\bm{x})$, and the kinetic energy $T$ can be written as 
\begin{equation}
    T=\frac{1}{2}a_j U_{jk} a_k - a_j W_{jk}b_k + \frac{1}{2}b_j V_{jk} b_k ,
\end{equation}
with the inertial tensors $U_{jk},~W_{jk}$ and $V_{jk}$, given below.
The variables $a_j$ and $b_j$ denote angular velocities of isospacial and spacial rotations, respectively, defined as $a_j=-i\Tr(\tau_jA^\dagger \pd_t A), ~b_j=i\Tr(\tau_j \pd_t BB^\dagger)$.
In the length scale $1/\kappa e$, one can write $M_\cl = \frac{\kappa}{e}M_\cl^*$ and $\left( U_{jk},  W_{jk},  V_{jk}\right)=\frac{1}{\kappa e^3}\left( U_{jk}^*,  W_{jk}^*,  V_{jk}^* \right) $ with the parameter-free quantities
\begin{align}
    & M_\cl^*=\int d^3x \left[\frac{1}{4}\Tr\left(\pd_k \n \pd_k \n \right)-\frac{1}{32}\Tr\left[\pd_j \n,\pd_k \n \right]^2 \right] ,
    \\
    & U_{jk}^*=\int d^3x\left[ 
    -\frac{1}{8}\Tr\left([\tau_j,\n][\tau_k,\n] \right)
    +\frac{1}{32}\Tr\left[[\tau_j,\n],\pd_l \n ][[\tau_k,\n],\pd_l \n \right] \right] ,
    \\
    &W_{jk}^*=i\int d^3x\left[ 
    \frac{1}{4}\Tr\left([\tau_j,\n]\rmL_k\n \right)
    -\frac{1}{16}\Tr\left[[\tau_j,\n],\pd_l \n ][\rmL_k\n,\pd_l \n \right] \right] ,
    \\
    &V_{jk}^*=\int d^3x\left[ 
    \frac{1}{2}\Tr\left(\rmL_j\n~\rmL_k\n \right)
    -\frac{1}{8}\Tr\left[\rmL_j\n,\pd_l \n ][\rmL_k\n,\pd_l \n \right] \right] ,
\end{align}
where $\rmL_j=-i\varepsilon_{jkl}x^k\pd_l$.

\subsection{
Angular momentum and the Hamiltonian}
The momenta conjugate to the collective coordinate $A$ and $B$ are defined by $\hat{K}_j=\frac{\pd L}{\pd a_j}$ and $\hat{L}_j=\frac{\pd L}{\pd b_j}$, respectively. The $\bm{\hat{K}}~(\bm{\hat{L}})$ is the body-fixed isospin (angular momentum) operator. The coordinate-fixed isospin $(\bm{\hat{I}})$ and spin $(\bm{\hat{J}})$ operators are defined through the orthogonal transformation
\begin{equation}
    \hat{I}_j=-R_{jk}(A)\hat{K}_k,\qquad
    \hat{J}_j=-R_{kj}(B)\hat{L}_k\, .
\end{equation}
These relations between the body-fixed and coordinate-fixed operators yield $\bm{\hat{K}}^2=\bm{\hat{I}}^2,~\bm{\hat{L}}^2=\bm{\hat{J}}^2$.
The Hamiltonian is defined through the Legendre transformation as 
\begin{equation}
    H=\hat{K}_ja_j+\hat{L}_jb_j-L.
\end{equation}

\subsection{
Axial symmetric ansatz}
For simplicity, we restrict ourselves to considering the axially symmetric Hopfions. It is well known that Hopfion solutions with topological charges $Q=1,2$ have an axial symmetry \cite{Gladikowski:1996mb,Battye:1998pe}.
To describe axially symmetric Hopfions, we employ an ansatz of the form
\begin{equation}
    \bm{n}=(\phi_1 \cos(\ell\theta) - \phi_2 \sin(\ell\theta), \phi_1 \sin(\ell\theta) + \phi_2 \cos(\ell\theta), \phi_3) ,
\end{equation}
where $l$ is an integer and $\bm{\phi}=(\phi_1,\phi_2,\phi_3)$ is a three-component unit vector depending only on $(\rho, z)$, with the cylindrical coordinates $(\rho, \theta, z)$.
The boundary condition for $\bm{\phi}$ is
\begin{equation}
    \bm{\phi}(0,z)=\bm{\phi}(\rho,z)|_{\rho^2+z^2\to\infty}=(0,0,1) .
\end{equation}
Due to the axial symmetry, all off-diagonal components of the inertia tensors vanish, except for the case where $\ell = \pm 1$, in which $W_{12} = \ell W_{21} \neq 0$.
The diagonal components hold the following relations:
\begin{align}
        &U_{11}=U_{22},~~V_{11}=V_{22}, \\
        &V_{33}=\ell W_{33}=\ell^2U_{33} .
        \label{eq:relation_inertia_tensors}
\end{align}
In addition, one finds that $U_{11}=U_{22}$ diverges under the boundary condition.
As a result, the Hamiltonian can be cast into a simple form
\begin{equation}
    H=M_\cl + \frac{1}{2V_{11}}\bm{\hat{L}}^2+\frac{1}{2}\left(\frac{1}{U_{33}}-\frac{\ell^2}{V_{11}} \right)\hat{K}_3^2 \ .
\end{equation}
The relation \eqref{eq:relation_inertia_tensors} leads to a constraint on the physical Hilbert space of the form
\begin{equation}
    \left(\hat{L}_3-\ell\hat{K}_3\right)\ket{\mathrm{phys}} =0 ,
    \label{eq:constraint}
\end{equation}
which gives a selection rule for possible glueball states.

\subsection{The energy eigenvalues and wavefunction}

Since this Hamiltonian has the same form as that of a symmetrical top, 
the energy eigenvalues can be given by
\begin{equation}
    E=\frac{\kappa}{e}M_\cl^* + \frac{\kappa e^3}{2}\left\{\frac{J(J+1)}{V_{11}^*}+\left(\frac{1}{U_{33}^*}-\frac{\ell^2}{V_{11}^*} \right)K_3^2 \right\} . 
\end{equation}
The eigenstate is given by a superposition of the basis $\ket{J,J_3,L_3}\ket{I,I_3,K_3}$ as
\begin{equation}
    \ket{\psi}=\displaystyle{\sum_{I\geq|K_3|}\sum_{I_3,J_3}}C^I_{I_3,J_3}\ket{J,J_3,L_3}\ket{I,I_3,K_3} ,
\end{equation} where $|J_3|,|L_3|\leq J$ and $|I_3|,|K_3|\leq I$, and $C_{I_3,J_3}$ are normalization constants.
In addition, the quantum numbers satisfy the constraint $L_3=\ell K_3$, which stems from Eq.~\eqref{eq:constraint}. 
Since the states $\ket{J,J_3,L_3}$ and $\ket{I,I_3,K_3}$ are eigenstates of a symmetric top, one can write 
\begin{equation}
    \braket{A}{I,I_3,K_3} \propto D^I_{I_3 K_3}(\alpha_1,\alpha_2,\alpha_3), \qquad
    \braket{B}{J,J_3,L_3} \propto D^J_{J_3 L_3}(\beta_1,\beta_2,\beta_3) ,
\end{equation}
with the Wigner $D$-functions, where 
 $\alpha_i$ and $\beta_i$ are Euler angles in isospace and real space, respectively
The wave function can be written as \begin{align}
    &\psi \equiv \braket{A,B}{\psi}
    \notag\\
    &=\sum_{I_3,J_3}C_{I_3,J_3}\braket{A}{I,I_3,K_3} \braket{B}{J,J_3,L_3}
    \notag\\
    &\propto \sum_{I\geq|K_3|}\sum_{I_3,J_3}C^I_{I_3,J_3}D^J_{J_3 L_3}(\beta_1,\beta_2,\beta_3)
     D^I_{I_3 K_3}(\alpha_1,\alpha_2,\alpha_3) \ .
\end{align}

\begin{table}[!t]
\centering
\caption{Classical masses and inertia tensors of Hopfions ($Q=\ell m$)
as dimensionless quantities.
}
\begin{tabular}{ccccc}
\hline\hline
$Q$ & $(\ell,m)$ & $M_\cl^*$ & $V_{11}^*$ &  $U_{33}^*$ \\
\hline
1  & (1,1) & 274.79  & 372.17   & 227.78  \\
2  & (2,1) & 447.20   & 968.33   & 267.72  \\
2  & (1,2) & 508.16  & 1203.1   & 398.09  \\ 
\hline\hline
\end{tabular}
\label{table:inertia_tensor}
\end{table}

\section{II.~Numerical procedure}
\label{sec:S2}

In this section, we explain the numerical procedure for 
constructing Hopfion solutions.
To describe axially symmetric Hopfions, we employ an ansatz of the form
\begin{equation}
    \bm{n}=(\phi_1 \cos(\ell\theta) - \phi_2 \sin(\ell\theta), \phi_1 \sin(\ell\theta) + \phi_2 \cos(\ell\theta), \phi_3) ,
\end{equation}
where $\ell$ is an integer and $\bm{\phi}=(\phi_1,\phi_2,\phi_3)$ is a three-component unit vector depending only on $(\rho, z)$, with the cylindrical coordinates $(\rho, \theta, z)$.
The boundary condition for $\bm{\phi}$ is
\begin{equation}
    \bm{\phi}(0,z)=\bm{\phi}(\rho,z)|_{\rho^2+z^2\to\infty}=(0,0,1) .
\end{equation}

To evaluate the classical mass and inertia tensors, we numerically construct Hopfion solutions.
We employ the following initial configuration appropriate to describing Hopfions of torus shape~\cite{Sutcliffe:2007ui}:
\begin{equation}
    \bm{\phi}=\frac{1}{1+F_1^2+F_2^2}(2F_1, -2F_2, F_1^2+F_2^2-1) ,
\end{equation}
where 
\begin{equation}
    F_1+iF_2 = \frac{\left(\cos f(\rho) + i\dfrac{z}{r}\sin f(\rho)\right)^m}{\left(\dfrac{\rho}{r}\sin f(\rho) \right)^\ell} ,
\end{equation}
with an integer $m$ and $r=\sqrt{\rho^2+z^2}$. The function $f(\rho)$ is chosen to satisfy the boundary condition $f(0)=\pi$ and $f(\infty)=0$.
The initial configuration possesses the Hopf charge $Q=\ell m$.
We numerically construct Hopfion solutions using a nonlinear conjugate gradient method with a fourth-order finite difference scheme.
The grid in the $\rho\text{-}z$ plane contains $75 \times 150$ points with a lattice spacing of $0.1$ in the length scale $1/\kappa e$.
The value of the classical mass and inertia tensors that the numerical solution possesses are listed in Table~\ref{table:inertia_tensor}.

\section{III.~Results with other 
parameter settings
}
\label{sec:S3}
\noindent

\begin{figure*}[!t]
    \centering
    \includegraphics[width=54em]{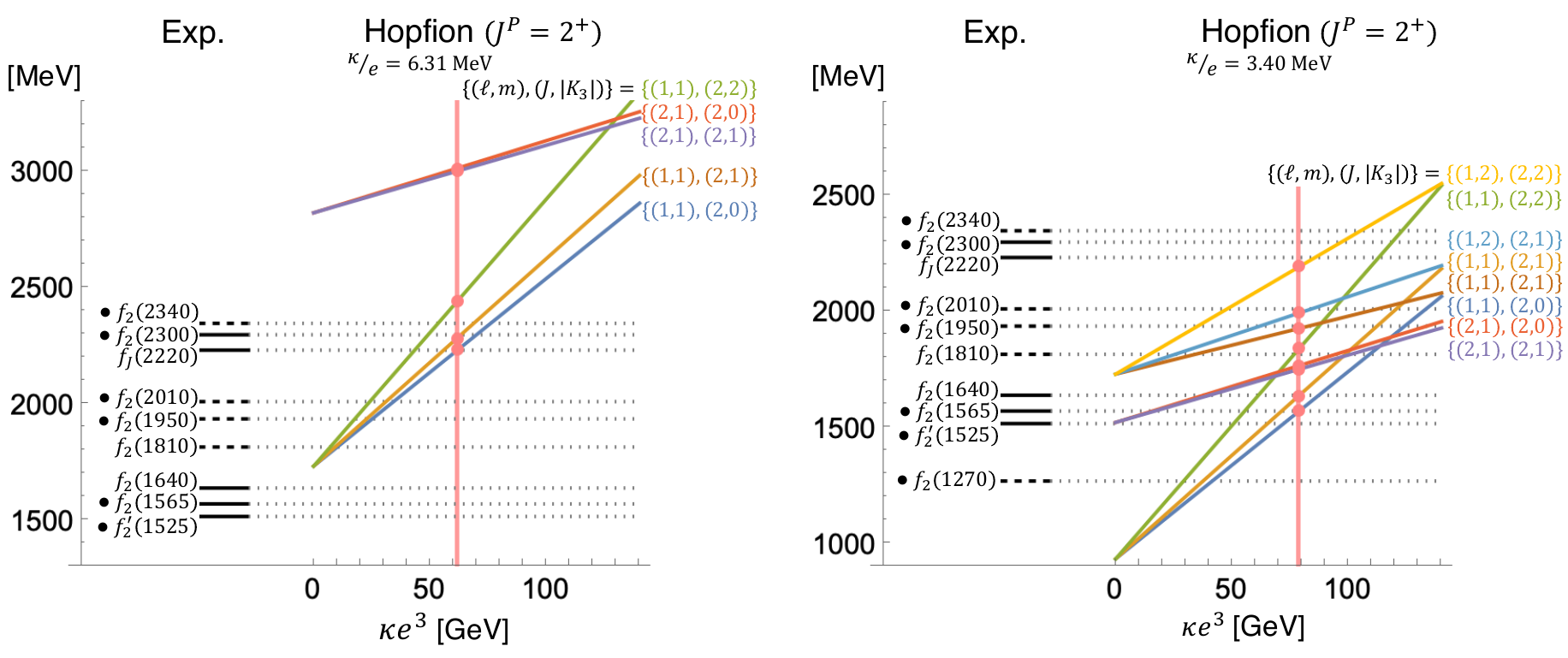}
    \caption{
    The energy spectra of the Hopfions with $J=2$
    for $\kappa/e=3.40$ MeV (Case 2; left panel) and $\kappa/e=6.31$ MeV (Case 3; right panel).
    By the red vertical dotted lines,
    the value of $\kappa e^3$ are indicated for which 
    the energy of
    the $Q=1$ Hopfion with
    $\{(1,1),(2,0)\}$
    is assigned
    to $f_2(1565)$  in Case 2
    and $f_{J=2}(2220)$ in Case 3.
    Notice that
    the
    $Q=1$
    Hopfion
    with $\{(1,1),(0,0)\}$
    is assigned to 
    $f_0(1500)$ in Case 2 and
    to $f_0(1710)$ in Case 3,
    as shown in Fig.~\ref{fig:f0}.}
    \label{fig:f2_supp}
\end{figure*}


\begin{figure}[!t]
    \centering
    \includegraphics[width=40em]{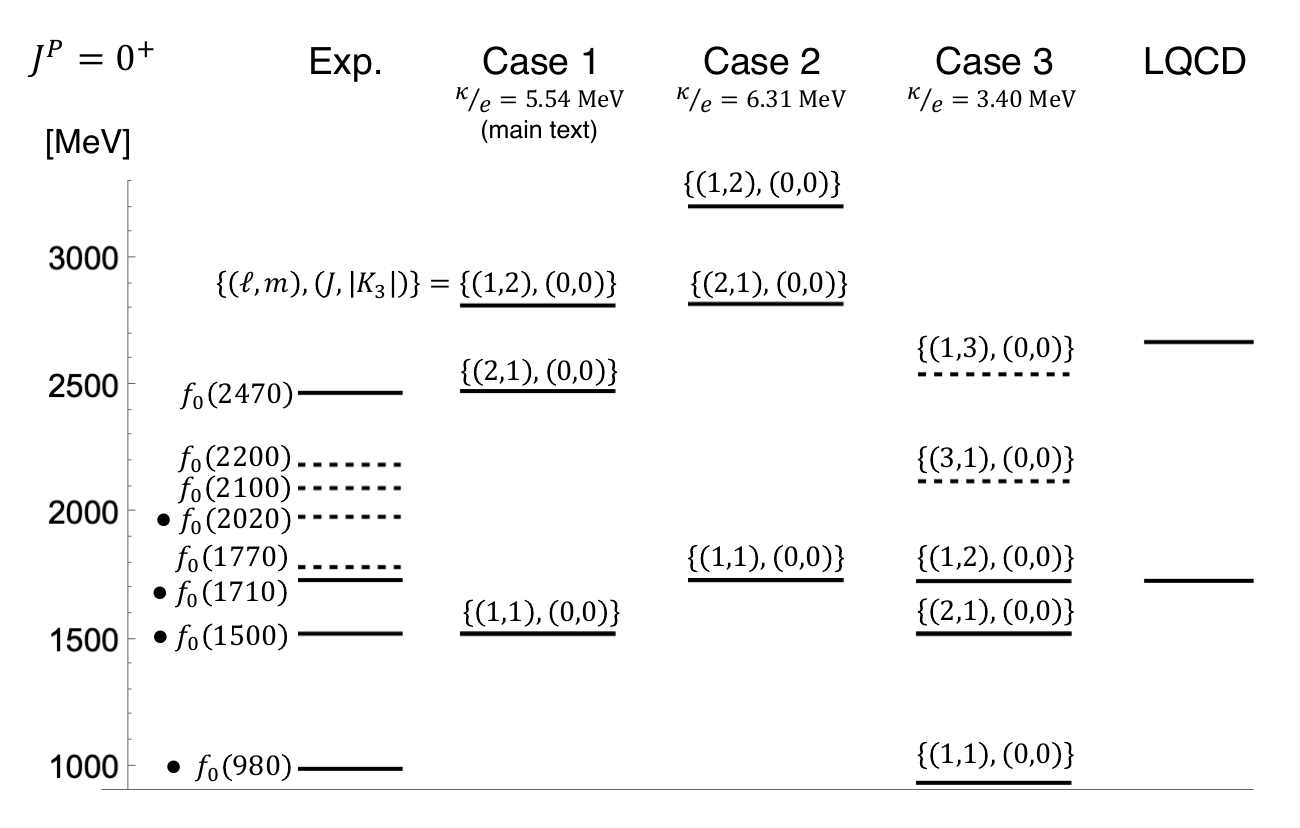}
    \caption{
    Comparison of Hopfion energy spectra
    {for $J^{P}=0^{+}$ case}
    with three different choices of coupling constants (Case 1, 2 and 3). 
    The states used to calibrate the parameters are represented in red. Hopfions with $(\ell,m)=(3,1)$ and $(1,3)$ are constructed by 3D full simulation, which is different from the others because they are not axially symmetric.
    On the l.h.s of the vertical axis, the mass spectra of light unflavored mesons collected in PDG~\cite{ParticleDataGroup:2022pth} is given.
    The LQCD data was taken from Ref.~\cite{Morningstar:1999rf}.
    The solid (dashed) line stands for particles with a decay width smaller (larger) than $150$ MeV, and the symbol $\bullet$ indicates established particles. 
    }
    \label{fig:spectrum_supp_0}
\end{figure}

\begin{figure}[!t]
    \centering
    \includegraphics[width=40em]{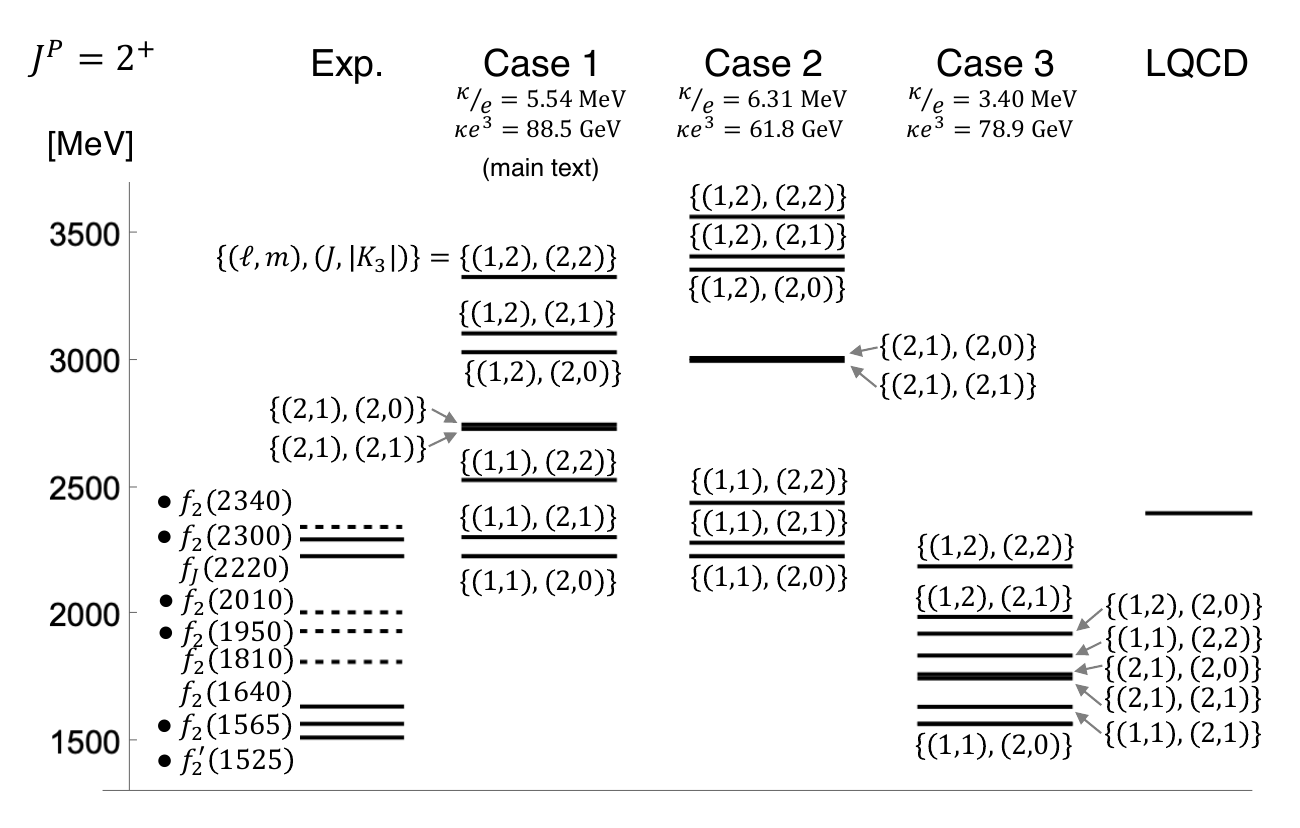}
    \caption{
    The $J^{P}=2^{+}$ case for Fig.~\ref{fig:spectrum_supp_0}.
    }
    \label{fig:spectrum_supp_2}
\end{figure}

\begin{table}[!t]
\centering
\caption{
Hopfion energies in unit of MeV for Cases 1, 2 and 3, see Eqs.~\eqref{eq:case1}-\eqref{eq:case3}.
The parameters are: $\kappa/e=5.54$ MeV and $\kappa e^{3}=88.5$ GeV in Case 1,
$\kappa/e=6.31$ MeV and $\kappa e^{3}=61.8$ GeV in Case 2,
and
$\kappa/e=3.40$ MeV and $\kappa e^{3}=78.9$ GeV in Case 3.
The Hopfions used for the fitting to experimental data are indicated by 
daggers.
}
\begin{tabular}{cccccc}
\hline\hline
$(\ell,m)$ & $(J,|K_3|)$   & & Case 1 & Case 2 & Case 3  \\
\hline
(1,1) & (0,0) & & \,\,1522$^{\dagger}$ & \,\,1733$^{\dagger}$ & \,\,935 \\
      & (2,0) & & \,\,2231$^{\dagger}$ & \,\,2231$^{\dagger}$ & \,\,1571$^{\dagger}$ \\
      & (2,1) & & 2306 & 2283 & 1638 \\
      & (2,2) & & 2531 & 2441 & 1840 \\
      \hline
(2,1) & (0,0) & & 2477 & 2820 & \,\,1522$^{\dagger}$ \\
      & (2,0) & & 2749 & 3011 & 1766 \\
      & (2,1) & & 2732 & 3000 & 1750 \\
      \hline
(1,2) & (0,0) & & 2814 & 3204 & 1729 \\
      & (2,0) & & 3034 & 3359 & 1926 \\
      & (2,1) & & 3108 & 3411 & 1992 \\
      & (2,2) & & 3330 & 3567 & 2191 \\
\hline\hline
\end{tabular}
\label{table:hopfion_energy_supp}
\end{table}

In the main text, we have considered the Hopfion energy spectra for the parameter sets in Eqs.~\eqref{eq:k/e} and \eqref{eq:ke3}.
We call this parameter set Case 1.
There, focusing on the importance of $f_{0}(1500)$ in $J^{P}=0^{+}$, we have considered it as the $Q=1$ Hofpion and set the parameter value $\kappa/e=5.54$ MeV in Eq.~\eqref{eq:k/e}, simultaneously reproducing the masses of $f_{0}(2430)$
as the $Q=2$ Hopfion.
We have also regarded $f_{J=2}(2220)$ in $J^{P}=2^{+}$ as the $Q=1$ Hopfion by setting the parameter value $\kappa e^{3}=88.5$ GeV in Eq.~\eqref{eq:ke3}, and
have found the triplet structure, i.e., $f_{J=2}(2220)$, $f_{2}(2300)$ and $f_{2}(2340)$.
However, one may consider that Case 1 is not the only possible parameter set.
Here, we investigate the other parameter settings and explore the possible energy spectra of Hopfions.

It was reported in the holographic QCD approach  that 
$f_{0}(1710)$ can be regarded as a glueball, while $f_{0}(1500)$ is unlikely to be~\cite{Brunner:2015oqa}.
Such case can be realized by the Hopfion approach by
setting 
\begin{equation}
\kappa/e=6.31 ~\text{MeV}    \mbox{ : Case 2} ,
\end{equation}
as shown 
at the grey vertical line on the right 
in Fig.~\ref{fig:f0}.
On the other hand, one can reproduce $f_{0}(1500)$ and $f_{0}(1710)$ simultaneously by setting 
\begin{equation}
\kappa/e=3.40 ~\text{MeV}    \mbox{ : Case 3} ,
\end{equation}
as shown 
at the grey vertical line on the left
in Fig.~\ref{fig:f0}.
In this case, however, it is inevitable 
for $f_{0}(980)$ to be  identified as a glueball, which 
is different from
the conventional understanding that $f_{0}(980)$ is an excited quark-antiquark state with a relative $P$-wave component (see, e.g., Refs.~\cite{Donoghue:1992dd,Amsler:2004ps}).
Indeed, this can be understood from the existence of an isotriplet scalar meson $a_{0}(980)$ as an isospin partner of $f_{0}(980)$.
In the present study, we 
call those parameter settings Cases 2 and 3
compared with Case 1, 
where
$f_{0}(1500)$ is excluded in Case 2 and $f_{0}(980)$ appears as a glueball in Case 3.
Nevertheless, those parameter settings exhibit interesting behaviors in the energy spectra of Hopfions.
The attempted parameter sets are summarized below:
\begin{align}
   &\text{Case 1:}
   \begin{array}{ll}
     &\{(1,1),(0,0)\} ~~\leftrightarrow~~ f_0(1500) \\
    &\{(1,1),(2,0)\} ~~\leftrightarrow~~ f_{J=2}(2220) 
   \end{array} 
   ~~\leftrightarrow~~ \kappa/e = 5.54\text{~MeV,~~} \kappa e^3 = 88.5\text{~GeV},
   \label{eq:case1} \\
   &\text{Case 2:}
   \begin{array}{ll}
     &\{(1,1),(0,0)\} ~~\leftrightarrow~~ f_0(1710) \\
    &\{(1,1),(2,0)\} ~~\leftrightarrow~~ f_{J=2}(2220) 
   \end{array} 
   ~~\leftrightarrow~~ \kappa/e = 6.31\text{~MeV,~~} \kappa e^3 = 61.8\text{~GeV},
    \label{eq:case2} \\
   &\text{Case 3:}
   \begin{array}{ll}
     &\{(2,1),(0,0)\} ~~\leftrightarrow~~ f_0(1500) \\
    &\{(1,1),(2,0)\} ~~\leftrightarrow~~ f_2(1565) 
   \end{array} 
   ~~~~~\leftrightarrow~~ \kappa/e = 3.40\text{~MeV,~~} \kappa e^3 = 78.9\text{~GeV}.
    \label{eq:case3}
\end{align}

Here, Case 1 discussed in the main text is shown for convenience in comparison to Cases 2 and 3.
We comment that $f_{2}(1430)$ and $f_{2}'(1525)$ are not taken into account as the candidate of glueballs despite of their smaller masses in our fittings, because they are considered to belong to radially excited quark-antiquark states with relative $P$-wave component ($1^{3}P_{2}$:  
principal number (node number) 1, total spin 1, $P$-wave and total angular momentum 2)~\cite{Amsler:2004ps}.

For  Cases 2 and 3, the dependence on $\kappa e^{3}$ in the Hopfion energy spectra are shown in Fig.~\ref{fig:f2_supp}.
Similarly to Case 1, the values of $\kappa e^{3}$ in Cases 2 and 3 are obtained by the fitting to the observed tensor mesons, as discussed in details below.

In Case 2,
we obtain $\kappa/e^{3}=61.8$ GeV by assigning the $Q=1$ Hopfion with $\{(1,1),(2,0)\}$ to $f_{J=2}(2220)$ in $J^{P}=2^{+}$, as indicated in the left panel of Fig.~\ref{fig:f2_supp}.
Here $f_{J=2}(2220)$ is chosen because it has the decay width $23^{+8}_{-7}$ MeV, relatively small number irrespective to its large mass.
This is the same assignment as Case 1 discussed in the main text. 
As a result, we obtain the Hopfion energy spectra in Figs.~\ref{fig:spectrum_supp_0} and \ref{fig:spectrum_supp_2} and the numerical values in Table~\ref{table:hopfion_energy_supp}.
It is interesting to see that $f_{2}(2300)$ and $f_{2}(2340)$ are comparable with the $Q=1$ Hopfions with $\{(1,1),(2,1)\}$ and $\{(1,1),(2,2)\}$, respectively, and hence $f_{J=2}(2220)$, $f_{2}(2300)$ and $f_{2}(2340)$ can be regarded as a triplet, as discussed in Case 1.
Similarly to Case 1, we also predict new tensor glueballs as the $Q=2$ doublet Hopfions with $\{(2,1),(2,1)\}$ and $\{(2,1),(2,0)\}$.

In Case 3, we
obtain $\kappa/e^{3}=78.9$ GeV by assigning
the $Q=1$ Hopfion with $\{(1,1),(2,0)\}$ to $f_{2}(1565)$ in $J^{P}=2^{+}$, as indicated in the right panel in Fig.~\ref{fig:f2_supp}.
In this fitting, we regard $f_{2}(1565)$, $f_{2}(1640)$ and $f_{2}(1810)$ as the triplet.
This assignment is reasonable because the ratio of the mass splittings among these tensor mesons,
$M_{f_{2}(1640)}-M_{f_{2}(1565)}:M_{f_{2}(1810)}-M_{f_{2}(1565)}=1:3.6$,
is close to the ratio of energy splitting of the  Hopfions, $1:4$, estimated from the mass formula in Eq.~\eqref{eq:q-energy}.
Note, however, that $f_{2}(1565)$, $f_{2}(1640)$ and $f_{2}(1810)$ have the decay widths $132\pm{23}$ MeV, $100^{+60}_{-40}$ MeV and $197\pm22$ MeV, respectively, the order of hundred MeV.

As summarized in Figs.~\ref{fig:spectrum_supp_0} and \ref{fig:spectrum_supp_2} and Table~\ref{table:hopfion_energy_supp}, in Case 3, there are many excited states in $J^{P}=0^{+}$ and $2^{+}$.
In $J^{P}=0^{+}$,
we find that the $Q=2$ Hopfions with $\{(2,1),(0,0)\}$ and $\{(1,2),(0,0)\}$ are comparable with $f_{0}(1500)$ and $f_{0}(1710)$, respectively.
We also find that the $Q=3$ Hopfions with $\{(3,1),(0,0)\}$ is close either to $f_{0}(2020)$, $f_{0}(2100)$ or $f_{0}(2200)$.
More interestingly, we find that the $Q=3$ Hopfion with $\{(1,3),(0,0)\}$ is near $f_{0}(2470)$, a narrow scalar meson, which has been identified to be $\{(2,1),(0,0)\}$ in Cases 1 and 2.
In $J^{P}=2^{+}$, similarly, the correspondence between the Hopfions and the tensor mesons is seen.
Similarly to Case 1 and 2, we find the doublet, i.e., the $Q=2$ Hopfion with $\{(2,1),(2,1)\}$ and $\{(2,1),(2,0)\}$ in Fig.~\ref{fig:f2_supp}.
It is particularly interesting that the $Q=2$ Hopfion with $\{(1,2),(2,2)\}$ is near $f_{J=2}(2220)$, a narrow tensor meson, which has been discussed in Cases 1 and 2.
Thus, $f_{J=2}(2220)$ is assigned to the triplet, i.e., $Q=2$ Hopfion with $\{(1,2),(2,0)\}$, $\{(1,2),(2,1)\}$ and $\{(1,2),(2,2)\}$, as presented in Fig.~\ref{fig:f2_supp}.
This assignment is different from that in Cases 1 and 2.


\end{document}